# Digital Twin conceptual framework for the O&M process of cubature building objects

Andrzej Szymon Borkowski, PhD Eng.

Warsaw University of Technology, Faculty of Geodesy and Cartography

ORCiD: 0000-0002-7013-670X

andrzej.borkowski@pw.edu.pl

**Abstract**: The broader construction industry is struggling with data loss, inefficient processes and low productivity in asset management. The remedy to these problems seems to be the idea of Digital Twin (DT). However, the frameworks proposed so far do not always support a solution to these problems. This paper conducts an extensive literature review to develop a conceptual framework for the Operation and Maintenance (O&M) phase for cubic facilities. The conceptual framework takes into account the increasingly popular Internet of Things (IoT) and Artificial Intelligence (AI) technologies. The presented framework, after appropriate modifications, can also be applied to infrastructure facilities or city fragments. The paper presents limitations and directions for further research. The DT paradigm has been adopted and its adoption is ongoing. Its implementation will progress in the coming years.

**Introduction**

The spread of Building Information Modeling (BIM) as a methodology for organizing building processes will gradually lead to digitized asset management. Taking it a step further, the Digital Twin (DT) concept is being used to integrate building assets with digital technologies to enable real-time analysis, as well as provide analysis and simulation capabilities. DT defined "A live digital coupling of the state of a physical asset or process to a virtual representation with a functional output" (Catapult, 2021), is potentially a universal definition as it is sector and domain independent (Boyes, Watson, 2022).The DT idea is expected to have a major impact on facilitating sustainable transformation through technological development (Schweigkofler et al. 2022). The maturity level of this technology will increase, and its spread will accelerate as its use increases in areas such as industry, health and smart city management (Erol, Mendi, Doğan, 2020). The DT paradigm can greatly benefit the built environment as a whole, but the lack of well-defined structural and functional descriptions limits the extent to which this technological paradigm can benefit the AECO - Architecture, Engineering, Construction and Operation (Delgado, Oyedele, 2021). DTs first appeared in manufacturing, where their role is to act as a detailed digital model that can be replicated in physical copies. In this case, the economic benefits come from detailed planning. In construction, DT techniques provide similar benefits, with the addition that information about the physical building landscape can be incorporated into planning. Facilities management, including building maintenance and asset management, also gains economic benefits from DT techniques, but requires periodic or continuous DT updates (Lehtola et al. 2022). The digital evolution is well explained by the CAD-BIM-DT juxtaposition (Fig. 1). Information systems have undergone a significant revolution from simple design-support applications, to increasingly sophisticated software for collecting data about a building site, to sophisticated database systems (Baghalzadeh Shishehgarkhaneh et al. 2022) that can use data from a variety of sources, including sensors from the so-called Internet of Things (IoT). BIM is primarily distinguished from CAD by its ability to add semantic non-graphical information. At the 3rd (highest) level of BIM maturity according to the Bew-Richards ramp, integration with sensors is expected, preferably in a telemetric manner (Esser, Vilgertshofer, Borrmann, 2023).

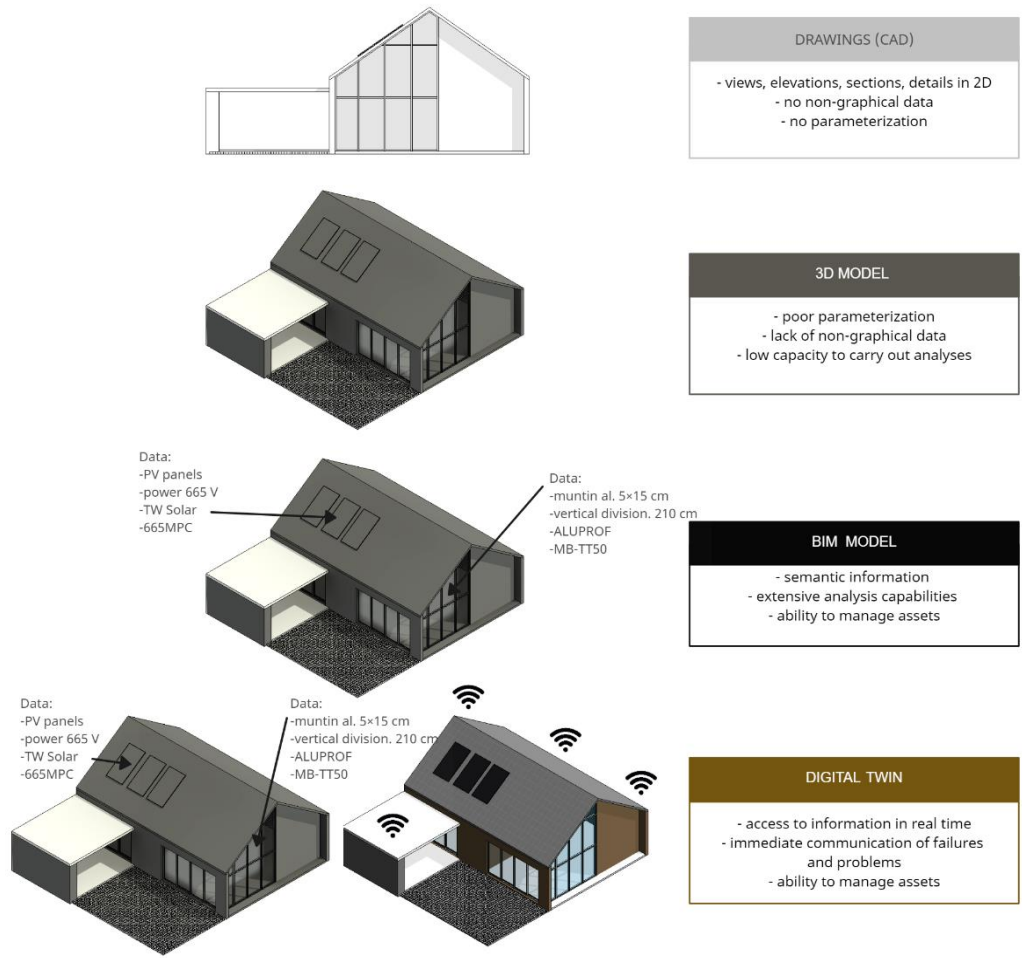

Figure 1: Evolution from CAD to the idea of the Digital Twin. Source: own elaboration.

The second half of the 20th century saw the continued development and use of CAD (Computer Aided Design) applications. Sprouting in the 1990s, the idea of BIM was just gaining popularity. Currently, more and more attempts are being made to integrate BIM with other technologies such as GIS (Geographic Information System), VR (Virtual Reality), AI (Artificial Intelligence) etc. to arrive at the idea of DT. In order to adapt BIM to newer, more integrated approaches at the micro (construction site) and macro (city neighborhoods) levels, a DT paradigm is required. The construction sector has already made significant progress since BIM's inception and has gained enough recognition and momentum to enable a shift from a static, closed information environment to a dynamic, network-based one that includes IoT integration and higher levels of AI implementation. This would help provide smarter building services, increased automation and consistency of information (Boje et al. 2020). DT-enhanced BIM can help drive the transition to full lifecycle and digital construction across the sector. Combining DT with a mature BIM framework can push BIM to further modernize and mutually introduce more unified and integrated DT applications in construction engineering (Honghong et al. 2023). Globally, the architecture and engineering (AEC) industry has seen an increase in the popularity of digital twin (DT) technologies due to their potential to improve collaboration and information communication throughout the project lifecycle, from design to operations and maintenance (O&M). However, empirical evidence for such adoption is fragmented, particularly for facility management (FM) activities during the O&M phase. The concept of DT has emerged relatively recently, and some problems still need to be solved. First, DT can easily be confused with BIM or Cyber Physical System (CPS), which are used in industry. Second, the components of DT applications in this sector are not well defined (Jiang et al. 2021). The frameworks defined so far focus only on selected aspects of O&M (Chen et al. 2021), or apply to buildings other than cubic structures (Ye et al. 2019; Mohammadi et al. 2023). Given this gap, a deep study of the literature and practical cases was conducted to develop a DT conceptual framework for the O&M phase. Research questions that can be posed in this context include: What solution will provide immediate access to information from various sources? Do these solutions need to be operated by people with high digital skills? Are there simpler solutions that can be operated by a browser and/or mobile devices?

**Methodology**

This paper uses several research methods: literature review, source criticism, abstraction, intuitive method, argumentation, and experience and reflection. Conceptual analysis was also applied to obtain basic information and process concepts that will define the future development of DT systems at the operation stage of cubic facilities. The article contributes to and extends the existing understanding of digital twins in the construction industry, viewing the idea of DT as an integral part of the transformation of facilities management (in the O&M phase) from reactive to proactive. To answer the research questions posed earlier, an in-depth literature review was conducted to understand the scenarios of DT application in FM of buildings in the O&M phase and to analyze the benefits and barriers to future implementation. It is considered useful to help this study demonstrate the empirical relevance of the theoretical proposition of the DT concept and its adoption in the real world. Three criteria are observed to limit the author's bias in selecting references. Two criteria were established to guide the review: (1) the freshness of the publications - the last few years; (2) the selection of theories and case studies that are representative and illustrative of various applications of DT, and can be useful in building a conceptual framework. After searching, selecting and studying relevant cases, different DT concepts and states of DT adoption were identified. While many examples illustrate the application of DT concepts, it should be emphasized that these cases, while representative and differing in characteristics, remain similar in their attempts to contribute to improving FM performance in the O&M phase through DT applications. Materials created by other researchers (i.e., secondary data) are increasingly available for reuse by the general research community (Fecher, Friesike, Hebing, 2015; Zhao et al. 2022). Most of the secondary data for this study, came from the scientific literature and some gray literature. Gray literature is defined as sources that have not been formally published in books and journals, but are found in technical reports, reprints, media and the like. It is gaining importance as a reliable source of data because of its value in disseminating scientific, technical, public and practical information (Søndergaard, Andersen, Hjørland, 2003). They oriented their attention to publications focusing on defining DT or its framework. For this reason, the typical bibliometric or systematic analysis was abandoned. In principle, studies using secondary data are common and as cost-effective as those using primary data. Therefore, the data from this article will provide a solid foundation for future theoretical research.

**Literature Review**

DT implementation is plagued by complexities such as interoperability, ineffective integration, inadequate information management, operational issues, lack of data in facilities management, and barriers to knowledge utilization and management throughout the lifecycle of a construction project (Ozturk, 2021). DT integrates artificial intelligence (AI), machine learning (ML) and data analysis to create dynamic digital models that are able to learn and update the status of the physical counterpart from multiple sources. Recent research results are helping to inspire innovative research ideas and promote widespread adoption of smart asset management using DT in the O&M phase (Lu et al. 2020). One study summarized several successful applications of DT from which conclusions and insights were drawn, and divided into three levels: (1) DT as just an enhanced version of the BIM model, (2) DT as semantic platforms for data flow, and (3) DT as AI-enabled agents for data analysis and feedback handling (Zhang et al. 2021). The evolution of BIM-based DTs represents a milestone for the strategic development of proactive planning and safety management solutions in the construction industry (Torrecilla-García, Pardo-Ferreira, Rubio-Romero, 2021). Deng et al. (2021) defined five levels of investigation from BIM to DT (Fig. 2). Level 1 is defined as a rich information model of a building object that is used not only in design, but also during construction and operation. Level 2 implies the use of BIM for analysis, simulation or reporting to support decision-making, including during the construction phase (Jiang, 2021). However, this requires a high level of digital skills that building owners and managers generally do not have. At level 3, there is integration with IoT, and this enables real-time monitoring and visualization of data. Level 4 is decision support using AI tools. This allows for quick, automatic or semi-automatic responses. The results of the research indicate that hybrid solutions and combining technologies (BIM, AI etc.) are helping to better support decision-making systems in construction projects (Sun, Liu, 2022). Level 5 involves additional human control, and in this case, applications or systems to support such work are necessary. In this way, there is interaction between the physical layer and the digital layer, with continuous human involvement (Klinc, Turk, 2019). The first two levels are being adopted by industry and enterprises (Wen et al. 2021), while the next two are in the adoption phase (Megahed, Hassan, 2022). The fifth level is the most challenging.

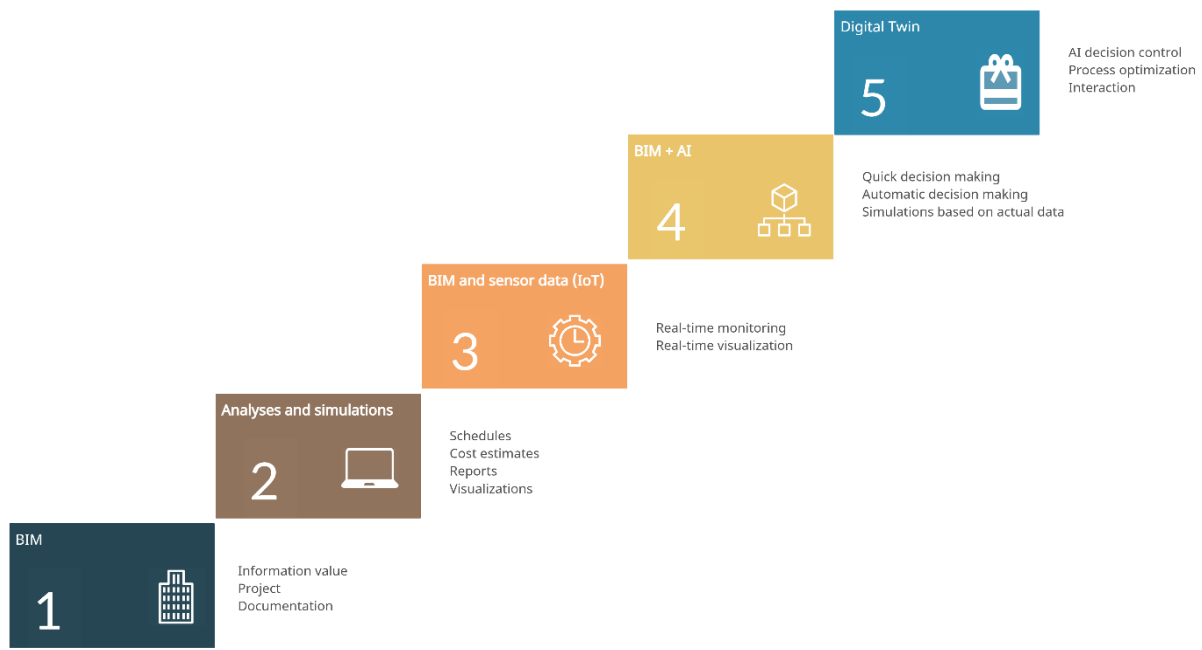

Figure 2. Levels of BIM development toward the Digital Twin. Source: own elaboration.

The authors of the study additionally show that the work grades well at Levels 2 and 3, which are BIM-supported simulation and BIM-IoT integration in the management of the built environment, more difficult challenges arise at Levels 4 and 5 (Deng, Menassa, Kamat, 2021). At Level 3, both location data and user feedback are combined with environmental data from sensors and spatial data from BIM data to generate the final resulting model (Abdelrahman, Chong, Miller, 2022). IoT devices are being deployed to collect real-time data on the current status of construction operations with little manual interaction. The rich source of IoT data serves as the basis for cyber-physical synchronization, which must be mapped to an IFC (or other) schema to ensure model interoperability, and then stored as event logs for data analysis and intelligent inference (Pan, Zhang, 2021). Data collection and analysis from specific sensors can be collected from the building's environment, façade or interior. BIM-IoT integration brings the creation of a more comprehensive building DT closer. This can be done using different types of sensors and communication protocols. Yet this doesn't change the fact that there are many technical obstacles to creating an ideal building DT (Khajavi et al. 2019). Along with a context-rich DT model, a network of IoT sensors is built to control environmental data, which is sent to a cloud server in real time. The data is processed to enrich the BIM model and enable facility managers to quickly understand environmental conditions (Wang et al. 2022).

Decision-making processes supported by DT and AI would greatly improve the intelligence and integration of the entire system. From standard aspects, effective and efficient communication, cooperation and management mode would be a close link between people and processes (Lu et al. 2020). Computational algorithms are used at various levels, and they also have many advantages to offer in DT for solving complex geometric challenges, such as optimizing production materials or reducing manual processes related to compliance checking. Computational algorithms effectively explore a large decision space and catch costly design/analytical errors in advance (Rausch et al. 2020). BIM is currently the cornerstone of DT construction, especially in the design, documentation, construction phases (Karmakar, Delhi, 2021). BIM models and related data are often stored, visualized and analyzed in the cloud (Afsari, Eastman, Shelden, 2016). Cloud and desktop solutions are the backbone of DT (Fig. 3).

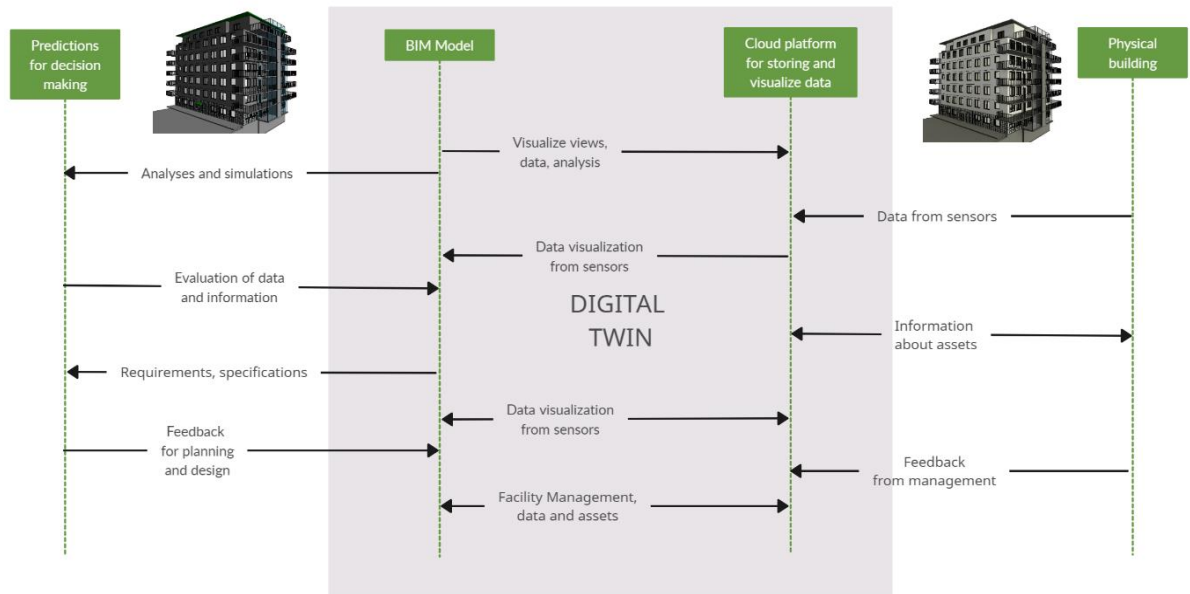

Figure 3. Example of an ideal Digital Twin for a cubature object. Source: own elaboration.

In the DT paradigm, cloud-based solutions are connected to sensors (Tan et al. 2022), in a physical building, to monitor various phenomena (temperature, gas and dust content, humidity, etc.). A virtual building in BIM transmits data on resources to manage them efficiently (Hagedorn et al. 2023). In all of this there is also a place for the metaverse or parts of it Virtual Reality (VR), Augmented Reality (AR) or Mixed Reality (MR) (Bale et al. 2022). All of this is intended to support cross-functional decision-making, which in turn supports learning. The main value of learning through a combination of predictive and prescriptive DT processes is to reduce downtime, failures, costs, energy waste and achieve the Sustainable Development Goals (SDGs). Such learning advantages are consistent with DT's capabilities for simulation, monitoring, life cycle assessment, detection, optimization and prediction (Sepasgozar, 2021).

**Results**

After reviewing the literature, it can be seen that although the concept of the digital twin was not explicitly mentioned in some publications, the titles and abstracts often referred to BIM-IoT or BIM-FM, and the content corresponds to the DT concept. To better understand the maturity level of DT in the construction industry, some frameworks are often used. Kritzinger et al. (2018) presented a classification of DT implementation in manufacturing. Their classification included three subcategories, each with a specific level of data integration. Digital Model is a digital physical representation that does not include any automated data exchange between physical and virtual spaces. Digital Shadow builds on the Digital Model subcategory and enables automated one-way data flow between the physical and virtual spaces. DT goes a step further, where physical and virtual space are fully integrated in both directions. Based on the research and the three subcategories of classification defined by (Kritzinger et al., 2018), it can be seen that the construction industry is moving beyond current BIM practices (El Jazzar, Piskernik, & Nassereddine, 2020), which typically focus on the use of digital models in design and construction. Much of the publication falls into the Digital Shadow subcategory, which in the context of the construction industry indicates that while data is collected and combined with the BIM model, changes made to digital models do not lead to changes in the physical space. For example, an MEP system can be monitored with a digital shadow; in the event of an emergency such as a leak, the digital model will indicate the problem but take no action. An ideal DT would not only shut down part of the MEP system, but also predict a potential failure before it occurs and suggest appropriate corrective measures.

To archive the full potential of DT, initiation should occur early in the project and throughout the facility lifecycle. Data collection should begin during the design phase using the BIM model. The data should then be continuously updated and collected throughout the life cycle of the construction facilities to achieve a fully functional as-built state, ready for the commissioning phase. During the operation and maintenance phase, the model aggregates data from various sensors. The data is stored and analyzed using cloud computing (i.e., data mining and big data). The virtual representation is then updated in real time with the necessary data and predictions about the physical object's behavior. This functionality gives the facility owner, manager or operator the ability to make informed decisions. Two-way communication between the physical and virtual facility also enables proactive maintenance. What's more, the long-term benefit of using this concept is to improve the next generation of construction projects using the knowledge captured in DT. Building the databases that will underpin DT requires the continuous integration of massive amounts of data throughout the lifecycle. From design, construction, the pre-O&M phase, to the O&M phase, where terabytes of static information and dynamic data can be integrated as a unified DT system, incorporating a building geometry model, attached property information, repair and maintenance systems, safety management systems and special facility business systems (Peng et al. 2020).

The conceptual framework of DT for the O&M phase for a cubic facility is presented below (Fig. 4). The basic assumption is the ability of management personnel to support advanced solutions - software as a service (SaaS) if they have adequate digital competence, or simpler solutions - Common Data Environment (CDE) services if they have basic competence. The foundation of DT is a BIM model developed on the basis of a physical model of a cubic object.

Sensor data through an appropriate communication protocol (e.g., MQTT) is sent to an operational database (cloud-based). The operational database can send data to a central database (cloud-based) or directly to the CDE. A BIM database that integrates data from the model and other sources can work in a similar way - send data to a central database or directly to a CDE. The CDE (it doesn't have to be one software, but a set) is the primary data repository for the O&M phase. It can collect data from the model, COBie, QRcodes, sensors, etc. Likewise with resources (devices, systems, equipment) that are relevant to managers. Data and resources can be classified, grouped or analyzed. The results of analysis or simulation can be visualized in the CDE environment. An important issue in this framework is the paradigm of two-way data exchange. For example, a failure detected by a sensor, triggers a decision by AI to shut down some part of the system, infrastructure or machine. The manager is immediately informed of the failure (e.g., by SMS, email, a message in the CDE) and can decide on an action (validate the AI's action, order an action or take their own action). Managers must also be able to view, update or edit the DT. This will support ongoing monitoring, control, supervision or medium-term modernization and renovation plans.

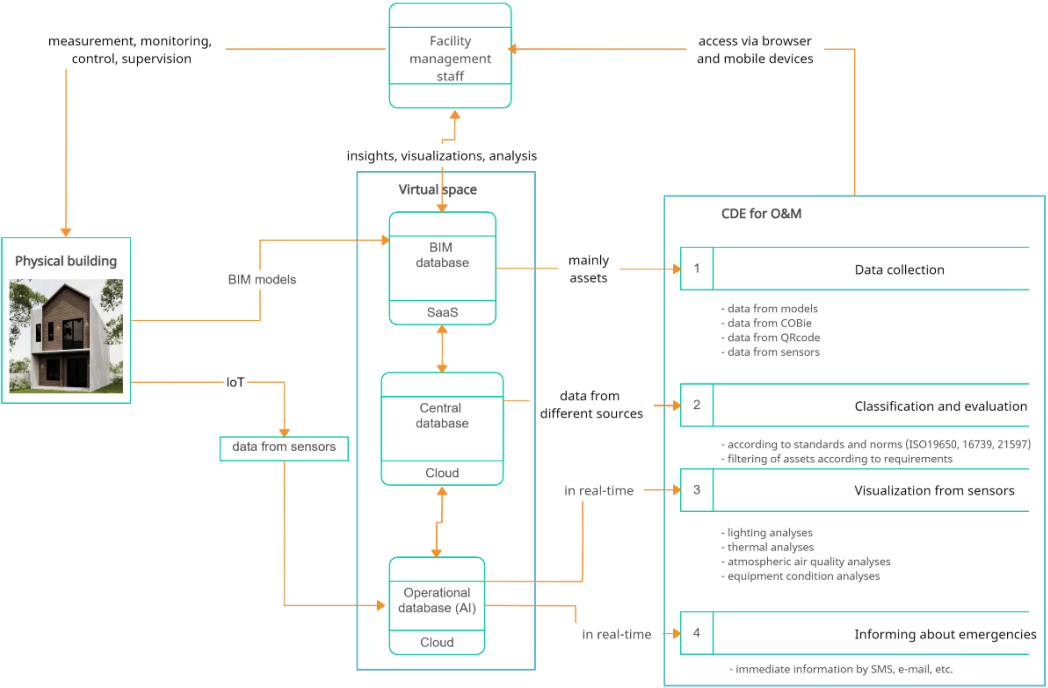

Figure 4. Digital Twin conceptual framework for a staff-managed cubicle facility. Source: own elaboration.

Despite the positive implications, there is a continued need for interdisciplinary collaboration with ICT engineers and AEC-FM practitioners to integrate IFC with the

RealEstateCore ontology to build seamless relationships between data from BIM models, CoBie sensors and IoT, streamlining data exchange standards for the industry (Arsiwala, Elghaish, Zoher, 2023). The research community is increasingly shifting its attention from BIM to DT applications. Although data interoperability has been extensively developed, resulting in several data exchange standards (e.g., IFC) and various computer implementations (e.g., Express, IFC, RDF, and OWL), the construction industry still faces data loss, mismatches due to data loss, mismatches and lack of semantics when importing and exporting data between BIM applications. Further research is still required to improve data interoperability, especially with the support of, for example, the DT or CPS data protocol (Almatared et al. 2022). Research suggests that AI would provide great potential to improve data acquisition and decision-making in the AEC industry. For example, AI could be used with IoT to pre-analyze raw data before sending it to DT. DT typically uses simulation to perform predictive analysis. However, AI can further improve predictive capability for better decision-making in AEC practice. In addition to data acquisition and predictive analysis, AI can also be used to support data exchange and interoperability.

The transition to a new era of digital information in the AEC-FM industry is directly related to DT. Based on the literature review and research results, it can be concluded that efforts have already been made to implement the DT concept in the AEC-FM industry. However, these efforts appear to be at a preliminary stage. Much research is needed to successfully add a fully functional DT model to the AEC-FM industry. In addition, there seems to be a parallel effort to modernize BIM to include through the implementation of Digital Twin in the AEC-FM industry. BIM seems to have the advantage of already being implemented for many assets, although there are challenges in integrating BIM and IoT and processing the data collected. DT has the advantage of having a good foundation for data processing and BIM integration. However, DT technology is even further behind in terms of research and implementation in the AEC-FM industry in the AEC-FM industry. Research on Digital Twin in the AEC-FM industry saw a significant increase in 2019 (Hosamo et al. 2022). Although problems are associated with a number of issues, such as data sharing limitations, design inadequacies and lack of a collaborative approach, throughout the lifecycle, the implementation and adoption of digital twins will grow. The shift toward BIM design appears to be already underway and unstoppable, due to the resulting economic and operational benefits (Doumbouya, Gao, Guan, 2016). Similarly, the use of DT in O&M processes may be seen as inevitable and certainly cost-effective in terms of efficiency and quality of services offered (Flamini et al. 2022). In the longer term, making BIM-DT models dynamic is undoubtedly an important step to qualitatively improve the human-machine interface without requiring excessive additional work, as the greatest effort is mainly related to the creation of BIM models, the economic return of which has been proven in many case studies. DT will soon become a widespread reality and will develop in all its potential.

**Limitations and future works**

The primary limitation was the availability of studies. The review was limited to papers written in a specific language (English), which largely excludes studies not reported in the literature published in other languages. Currently, BIM and DT are dominated by English, resulting, among other things, in an overrepresentation of studies conducted in Anglo-Saxon countries. This leads to problems in generalizing the results (external validity) to other populations, interventions or socio-institutional contexts - other than those covered by the studies analyzed. Another limitation was the quality of available research. Many theories in scientific publications may not meet quality criteria, may lack the basic information necessary to replicate the study, or may represent the authors' subjective experiences. A universal problem is the burden of selective publication of research results (publication bias). Scientific articles typically do not faithfully reflect all of the research and analysis conducted. Although the study had a limited sample of sources, the data collected from them were subject to bibliometric limitations. In addition, only academic research is used in scientometric mapping and analysis. This means that practical and commercial innovations were somewhat excluded. To collect better future research, data from practitioners and companies can be used.

Future research should include a comprehensive viewpoint to deal with the difficulties mentioned in this study. AECO industry stakeholders and researchers will benefit from the results of this research, which can broaden awareness of current research goals, research gaps, and long- and short-term future research trends in DT research. Moreover, the study did not examine the organizational effectiveness of the implementation of the digital twin for asset management, which is another area for future research to motivate various project participants to adopt it and increase data transparency, breaking down industry conundrums. In the future, empirical evaluation of the proposed framework through larger-scale implementation and demonstration to industry practitioners will further validate, explore and improve its wider application in cubic facilities, perhaps eliminating potential challenges.

**Conclusion**

The DT conceptual framework for O&M presented in the paper can be put into practice through application in a cubic facility. The results of such an experiment could bring closer the real-world adoption of DT in enterprises and later in national economies. The shift in emphasis from BIM to DT is becoming increasingly apparent in the scientific world. With appropriate modifications, the framework can also be used for an infrastructure facility or housing development. More and more spaces are being digitized, and the cyber-physical connection is increasingly desirable. Adoption of DT seems inevitable, although much more effort is needed from practitioners and researchers in this direction. The advantages and benefits are certainly more numerous than the barriers and risks posed by the use of DT.